\documentclass[runningheads]{llncs}

\usepackage[T1]{fontenc}
\usepackage{graphicx}
\usepackage{multirow}
\usepackage{balance} 
\usepackage{stfloats}
\usepackage{xcolor}
\usepackage{booktabs}
\usepackage{xspace}
\usepackage[normalem]{ulem}
\usepackage{amssymb}
\usepackage{url}
\usepackage{amsmath}

\newcommand{\toolname}[0]{\textsc{VIC-RAGENT}\xspace}

\begin{document}

\title{Detecting Vulnerability-Inducing Commits via Multi-Stage Reasoning with LLM-Based Agents}
\titlerunning{Detecting VICs via Multi-Stage Reasoning with LLM-Based Agents}

\author{
Liyou Chen\inst{1}
\and
Hailong Sun\inst{1,2,}\thanks{Corresponding authors}
\and
Xiang Gao\inst{1,2,\star}
\and
Yue Pan\inst{3}
}

\authorrunning{Chen et al.}

\institute{
State Key Laboratory of Complex \& Critical Software Environment (CCSE), Beihang University, Beijing, China.
\email{chenliyou@buaa.edu.cn}
\and
Hangzhou Innovation Institute of Beihang University, Hangzhou, Zhejiang, China.
\email{sunhl@buaa.edu.cn},
\email{xiang\_gao@buaa.edu.cn}
\and
North China Municipal Engineering Design \& Research Institute Co., Ltd., Beijing, China.
\email{loveangel\_3344@yeah.net}
}

\maketitle

\begin{abstract}
Detecting vulnerability-inducing commits (VICs) at submission time is critical for improving the security and reliability of software systems.
However, this task is highly challenging because it requires reasoning about the semantic impact of code changes from heterogeneous information sources, including code diffs, commit messages, and the surrounding contextual code.
Existing approaches often struggle to fully capture these complex interactions, resulting in limited detection performance.
In this paper, we propose \toolname, an LLM-based multi-agent framework for effective and explainable vulnerability detection.
\toolname leverages multiple specialized agents to provide complementary perspectives, including structural analysis, intent understanding, and vulnerability inspection.
To further improve detection reliability, the framework employs a multi-stage reasoning process that progressively refines candidate vulnerabilities through preliminary inspection, reanalysis, and a final decision stage.
Experimental results on a real-world dataset across multiple LLMs demonstrate that \toolname consistently outperforms baselines, including Direct, CoT, and CodeAgent.
Compared to the strongest baseline, \toolname achieves 1.2–1.7× higher F1-scores across different models.
Overall, \toolname offers a robust, explainable, and practical solution for detecting VICs in modern software development workflows.

\end{abstract}

\section{Introduction}
\label{sec:intro}

Open source software (OSS) has become fundamental infrastructure for modern software systems~\cite{oriol2023comprehensive,CHEN2026113001}, but its collaborative development model enlarges the software supply chain attack surface~\cite{shen2025understanding}.
In large-scale projects with continuous integration, even minor code changes may unintentionally introduce exploitable weaknesses~\cite{jiang2024understanding}.
Prior studies~\cite{6747164,jiang2024understanding} show that many vulnerabilities originate from vulnerability-inducing commits (VICs)~\cite{bao2022v}, which introduce security flaws during routine development activities such as feature additions and refactoring~\cite{iannone2022secret,woo2025large}.

Detecting whether a newly submitted commit introduces a vulnerability---known as Just-In-Time Vulnerability Detection (JIT-VD)~\cite{nguyen2024code}---remains a challenging task.
JIT-VD aims to detect whether a commit introduces vulnerabilities based on code changes and contextual information.
Existing approaches include feature-based methods and deep learning models~\cite{pornprasit2021jitline,yang2017vuldigger}, but they struggle to capture commit semantics and developer intent.
To address these limitations, deep learning approaches have been proposed to learn representations from code and diffs, including neural vulnerability detection systems~\cite{li2018vuldeepecker,li2021vuldeelocator}, graph-based models~\cite{zhou2019devign,cao2021bgnn4vd}, and transformer-based models~\cite{fu2022linevul}.
More recently, JIT-specific models~\cite{nguyen2024code,sun2025hgtjit} attempt to capture the semantics of code changes.

Recent advances in LLMs have demonstrated strong capabilities in code understanding and reasoning~\cite{wang2025contemporary,ni2026learning,yang2025context}.
LLM-based approaches have been applied to vulnerability detection and software security analysis~\cite{lu2024grace,zhou2025large,tian2025enhanced}, including prompt-based methods and retrieval-augmented methods~\cite{du2024vul}.
In addition, agent-based frameworks such as CodeAgent~\cite{tang2024codeagent} explore automated code review.
However, these approaches are primarily designed for general vulnerability detection or code review tasks, and lack specialized mechanisms for reasoning about VICs in evolving codebases.
The ReAct Agent~\cite{yildiz2025benchmarking} introduces an iterative reasoning process based on a thought--action--observation loop, enabling LLMs to dynamically acquire additional context during analysis.
This design may lead to the accumulation and propagation of errors across reasoning steps, while lacking explicit mechanisms for result verification.
Such a design may make intermediate results difficult to independently verify and limit the transparency of the reasoning process.
In contrast, role-specialized multi-agent systems have been shown to improve modularity and reasoning reliability by explicitly separating different reasoning objectives~\cite{motwani2024malt,2024Asurveyonllm}.

To address these limitations, we propose \toolname, an LLM-based multi-agent framework with a multi-stage reasoning process for detecting of VICs.
\toolname introduces a structured multi-stage reasoning pipeline with role-specialized components, enabling explicit decomposition of commit semantics, intent understanding, and vulnerability verification.
\toolname analyzes each commit by integrating multiple sources of information, including code diffs, commit messages, and file-level context.
The framework adopts a multi-agent analysis workflow.
Specifically, a \emph{Code Analyst} agent extracts structural information, and a \emph{Target Analyst} agent interprets commit intent.
\emph{Vulnerability Inspector} agents perform multi-stage 
reasoning to identify potential vulnerabilities.
When a vulnerability is confirmed, a \emph{Document Specialist} agent generates a security report that is stored in the knowledge base, enabling future analyzes to leverage historical vulnerability cases.

We evaluate \toolname on the V-SZZ dataset~\cite{bao2022v} across multiple LLMs.
Experiments show that \toolname consistently improves recall and F1-score across all evaluated LLMs, achieving up to 2× recall improvement over the strongest baseline.

Our main contributions are summarized as follows:

\textbf{A multi-agent framework for JIT-VD.}
We propose \toolname, a multi-agent framework that analyzes potential vulnerabilities introduced by commit.

\textbf{A multi-stage reasoning process.}
We introduce a structured inspection workflow consisting of preliminary analysis, reanalysis, and final decision, enabling coarse-to-fine reasoning for improved vulnerability detection.

\textbf{Knowledge-augmented vulnerability reasoning.}
We design a vulnerability knowledge base that stores security reports organized by vulnerability types, allowing \toolname to leverage historical vulnerability cases through retrieval-augmented reasoning.

\textbf{Empirical evaluation on a real-world dataset.}
We conduct experiments on a real-world dataset,
demonstrating strong detection performance across three LLMs.

The implementation of \toolname and the dataset are available at: \url{https://github.com/KeLeXueBi/VIC-RAGRENT}.
\section{Multi-Agent Analysis Framework}
\label{sec:multi-agent analysis framework}

\subsection{Framework Overview}
\label{sec:framework overview}

\begin{figure}[!tbp]
  \centering
  \includegraphics[width=\linewidth]{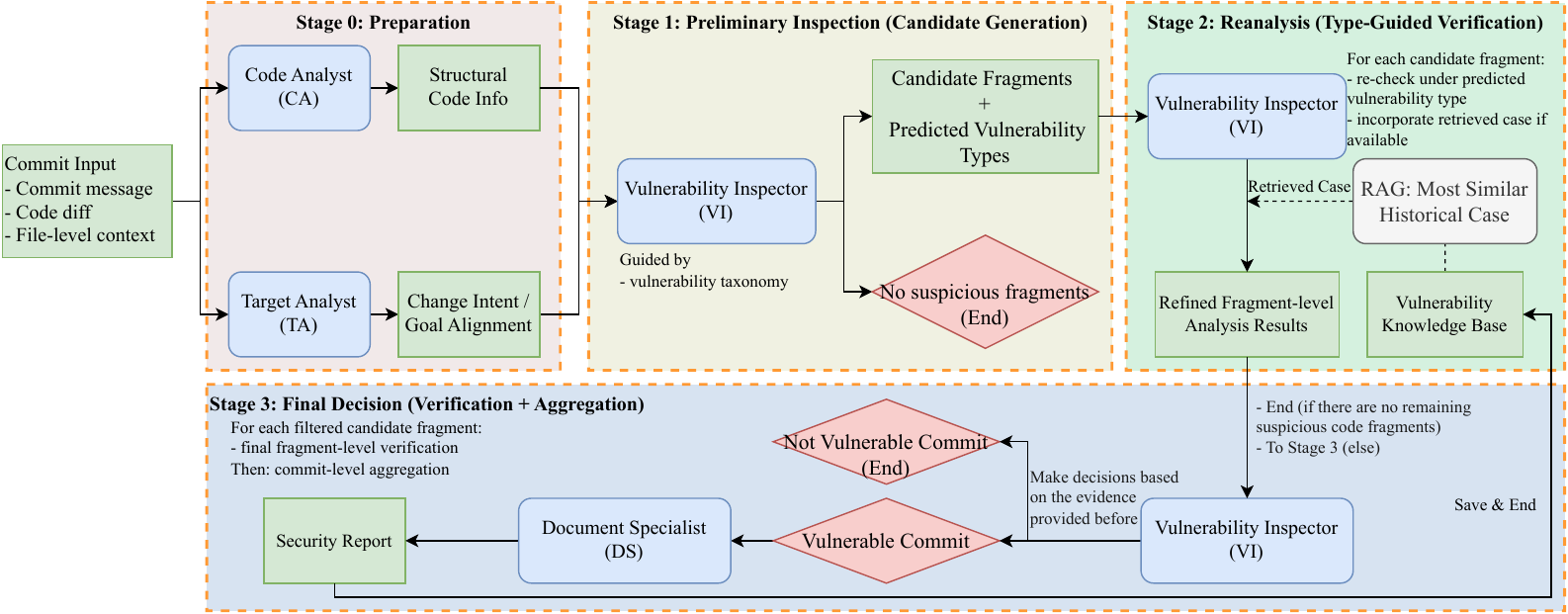}
  \caption{The overall framework of \toolname.}
  \label{fig:framework}
\end{figure}

\begin{table}[!tbp]
    \centering
    \caption{Agents in \toolname.}
    \label{tab:agents}
    \small
    \begin{tabular}{p{3.3cm}|p{2.3cm}|p{3.1cm}|p{3.3cm}}
        \toprule
        \textbf{Agent} & \textbf{Input} & \textbf{Output} & \textbf{Responsibility} \\
        \midrule
        Code Analyst (CA) & Diff + Context & Structural summary & Structural analysis \\\midrule
        Target Analyst (TA) & Diff + Message & Intent summary & Intent understanding \\\midrule
        Vulnerability Inspector (VI)(stage 1) & CA + TA & Candidate vulnerabilities & Security inspection \\\midrule
        Vulnerability Inspector (VI)(stage 2) & VI (stage 1) + Knowledge Base & Refined vulnerabilities & Knowledge-augmented verification \\\midrule
        Vulnerability Inspector (VI)(stage 3) & VI (stage 2) & Prediction result & Final decision \\\midrule
        Document Specialist (DS) & Verified VIC & Security report & Knowledge construction \\\midrule
        Audit Supervisor (AS) & Agent outputs & Validation feedback & Process verification \\
        \bottomrule
    \end{tabular}
\end{table}

Fig.~\ref{fig:framework} illustrates the overall architecture of \toolname, a multi-stage reasoning framework for detecting VICs.
Table~\ref{tab:agents} summarizes the responsibilities of each agent in the framework.
\toolname performs a structured analysis through four sequential stages: preparation, preliminary inspection, reanalysis, and final decision.
The output of each stage is progressively refined and passed to subsequent stages.

\subsection{Stage 0: Preparation}
\label{sec:stage0}

Given a commit input, the preparation stage performs contextual analysis using two agents.
The goal of this stage is to construct structured intermediate representations that capture both structural and semantic aspects of the code changes.
The structural representation is constructed by the Code Analyst (CA), which extracts structural information from commit changes to help other agents understand how modified code interacts with the existing system.
Given the code diff together with the file-level context, the CA summarizes structural relationships among components, including simplified call graphs and potential data flow patterns.

The intent representation is derived by the Target Analyst (TA), which interprets the intentions behind commit changes and aligns the goals described in the commit message with the modified files.
The TA extracts the goals described in the commit message and categorizes them into common modification types (e.g., bug fixes, feature additions, refactoring, performance optimizations, or security patches).
Inconsistencies between intended goals and implemented changes—such as incomplete implementation or unclear scope—may indicate potential risks.

Together, the structural representation and intent representation provide complementary structural and semantic views of the commit.
This dual-view representation reduces ambiguity in subsequent reasoning stages and helps the Vulnerability Inspector better interpret the intent and impact of code changes.

\subsection{Stage 1: Preliminary Inspection}
\label{sec:stage1}

The preliminary inspection stage can be viewed as a high-recall candidate generation step that approximates an over-complete set of potential vulnerability hypotheses.
Instead of making definitive decisions, this stage intentionally relaxes verification constraints and constructs a superset of plausible vulnerability-inducing fragments.

Based on the structural information extracted by CA and the change intentions identified by TA, the Vulnerability Inspector (VI) examines modified code fragments and analyzes potential vulnerability-inducing patterns across several vulnerability categories.
The analysis is not limited to the modified lines; when a change affects function behavior, variable usage, or call flow, the inspector further examines relevant surrounding code to assess potential risks.

The inspection follows a category-driven analysis strategy based on common vulnerability types observed in real-world systems.
Specifically, VI evaluates code changes with respect to six vulnerability types: I/O Validation, Memory Safety, Web Security, Authentication and Authorization, Resource Management, and File/Path Handling.
If suspicious code fragments are identified, VI records the relevant code segments together with their associated vulnerability type and an explanation of the potential risk.
These preliminary findings are then passed to the next inspection stage for further verification and refinement.
This design is particularly important for VIC detection, where vulnerabilities are often subtle and may not exhibit explicit security violations at the time of introduction.
By allowing the model to flag potentially risky patterns with incomplete evidence, this stage prioritizes recall and ensures that suspicious candidates are not prematurely discarded.

\subsection{Stage 2: Reanalysis}
\label{sec:stage2}

While the preliminary inspection may produce uncertain or weakly supported candidates, this stage refines them through constrained reasoning over the hypothesis space.
Each hypothesis is re-evaluated under additional constraints, including its predicted vulnerability type and contextual consistency with the commit.
Compared to Stage 1, which prioritizes coverage, this stage focuses on improving the reliability of hypotheses by enforcing structured constraints.

In practice, the Vulnerability Inspector (VI) performs type-guided reanalysis on each candidate fragment based on its predicted vulnerability category.
In addition, the system may retrieve similar historical cases from the vulnerability knowledge base using a retrieval-augmented generation (RAG) mechanism.
When available, similar historical cases are retrieved from the vulnerability knowledge base and incorporated as additional evidence.
If no sufficiently similar cases are found, the analysis proceeds using only the available contextual information.
This design enables case-based reasoning while avoiding over-dependence on the knowledge base.
The output of this stage is a refined set of fragment-level hypotheses, which are passed to the final decision stage.

\subsection{Stage 3: Final Decision}
\label{sec:stage3}

Although the reanalysis stage refines candidate hypotheses, uncertainty in LLM reasoning may still affect the final assessment.
This stage produces a commit-level vulnerability decision through structured verification and aggregation.

Given the refined hypothesis set, each hypothesis is independently verified under a more conservative reasoning setting.
A global decision is then derived by aggregating fragment-level results.

In practice, VI performs fragment-level verification under a low-temperature configuration to produce more deterministic and conservative judgments.
A commit is classified as vulnerability-inducing if at least one fragment is confirmed as vulnerable; otherwise, it is classified as non-vulnerable.

If the commit is determined to be vulnerable, the Document Specialist (DS) generates a security report summarizing the detected issues, which is stored in the vulnerability knowledge base for future retrieval.
Throughout the process, the Audit Supervisor (AS) monitors agent outputs and verifies whether they satisfy predefined requirements, ensuring consistency of the multi-stage reasoning process.

\subsection{Vulnerability Knowledge Base}
\label{sec:vulnerability knowledge base}

To support knowledge-augmented vulnerability reasoning, \toolname maintains a vulnerability knowledge base that stores historical vulnerability analysis results generated during previous inspections.
This knowledge base enables the system to reuse prior vulnerability cases during the reanalysis stage.
The knowledge base is organized using a two-level structure.
At the top level, reports are grouped by high-level vulnerability categories, and at the second level, each entry contains a security report describing a VIC.
To enable efficient retrieval, \toolname maintains a vector index over the stored cases.
Each commit is encoded into a fixed-length embedding using a pretrained code model (e.g., CodeBERT~\cite{feng2020codebert}), and the embeddings are associated with the corresponding reports for similarity-based retrieval.

During the reanalysis, the system retrieves similar historical cases based on embedding similarity.
The retrieved cases provide reference examples that support case-based reasoning, helping the model assess whether the current commit exhibits similar vulnerability patterns.
Whenever a commit is confirmed to be vulnerable by \toolname, its security report and embedding are added to the knowledge base, enabling continuous expansion.
To mitigate this issue, the system periodically removes reports corresponding to FP cases, retaining only reliable vulnerability instances and preventing error propagation. 
\section{Implementation}
\label{sec:implementation}

\subsection{LLM Configuration}
\label{sec:llm configuration}

\begin{table}[!tbp]
    \centering
    \caption{Temperature configuration for different agent roles / Stages.}
    \label{tab:temperature_config}
    \small
    \begin{tabular}{l c}
        \toprule
        \textbf{Role / Stage  } & \textbf{Temperature} \\
        \midrule
        CA & 0.2 \\
        TA & 0.2 \\
        VI (Stage 1) & 0.4 \\
        VI (Stage 2) & 0.4 \\
        VI (Stage 3) & 0.1 \\
        DS & 0.3 \\
        AS & 0.1 \\
        \bottomrule
    \end{tabular}
\end{table}

\toolname relies on LLMs to perform reasoning tasks across different agent roles.
We evaluate the framework using three LLMs: DeepSeek-V3.2~\cite{liu2024deepseek}, Qwen-Plus~\cite{bai2023qwen}, and GPT-4o-mini~\cite{openai2024gpt4omini}.
Each agent role is implemented with the same LLM per experiment.
\toolname is evaluated using three LLMs: DeepSeek-V3.2~\cite{liu2024deepseek}, Qwen-Plus~\cite{bai2023qwen}, and GPT-4o-mini~\cite{openai2024gpt4omini}.
All agent roles use the same underlying LLM within each experiment.
Table~\ref{tab:temperature_config} summarizes the temperature settings adopted for different stages and agent roles.
The same prompts and workflows are used across all LLMs.
To ensure a fair comparison, all experiments use the same prompts and agent workflows across all LLMs.

\subsection{Agent Prompt Design}

\toolname uses role-specific prompt templates with JSON outputs for intermediate communication; full prompts and implementation details are publicly available.

\subsection{Embedding and Retrieval}
\label{sec:embedding and retrieval}

Commit diffs are encoded using CodeBERT~\cite{feng2020codebert} and stored in a vector index.
During Stage 2, the current commit is encoded with the same model and matched against historical cases using cosine similarity.
Only the most similar case is retrieved and used when similarity exceeds 0.85.

\textbf{Knowledge base maintenance.}
Since the knowledge base is incrementally updated using model-generated reports, it may be affected by false positive (FP) predictions.
To mitigate this issue, we periodically remove reports corresponding to FP cases, retaining only reliable vulnerability instances.
For example, if a commit is predicted as vulnerable by the preliminary analysis but the ground truth label indicates it is not, it is considered an FP and removed.
\section{Experimental Setup}
\label{sec:experimental setup}

\subsection{Research Questions}
\label{sec:RQ}

To evaluate the effectiveness of \toolname for detecting VICs, we design a series of experiments to answer the following research questions.

\textbf{RQ1: How effective is \toolname in detecting VICs compared with existing approaches?}

\textbf{RQ2: How does the multi-agent collaboration contribute to vulnerability detection performance?}  

\textbf{RQ3: What is the computational cost of \toolname, and how does it compare with baseline methods?}

\textbf{RQ4: Can \toolname generalize to unseen vulnerabilities without relying on memorized knowledge?}

\subsection{Dataset}
\label{sec:dataset}

We evaluate \toolname on the V-SZZ dataset~\cite{bao2022v}, which contains manually curated vulnerability-inducing commits (VICs) and vulnerability-fixing commits (VFCs).
Following the original evaluation protocol, we start from 360 commits (172 VICs and 188 VFCs).

To construct a binary classification dataset for VIC detection, commits labeled as both VIC and VFC are treated as VICs.
We further remove commits that exceed the LLM context window, cannot be decoded, cannot be retrieved from the original repositories, or fail during framework execution.

After preprocessing, the final dataset contains 241 commits, including 106 VICs and 135 VFCs.
Among these commits, approximately 80\% belong to paired vulnerability records, meaning that a VFC is associated with one or more corresponding VICs.

\subsection{Baselines}

We compare \toolname with three representative baselines based on LLMs.

\textbf{Direct LLM.}
In this baseline, the commit information (including the commit message, code diff, and file context) is directly provided to the LLM, which produces a binary prediction indicating whether the commit introduces a vulnerability. 
This setting represents a vanilla prompting approach without structured reasoning or agent-based decomposition.

\textbf{Chain-of-Thought (CoT).}
This baseline extends direct prompting by instructing a single LLM to perform explicit step-by-step reasoning before producing the final binary decision. 
The model is asked to summarize the commit changes, identify security-relevant code fragments, analyze whether the changes may introduce a new vulnerability, and then output a commit-level yes/no prediction. 

\textbf{CodeAgent.}
CodeAgent~\cite{tang2024codeagent} is an LLM-based agent framework designed for automated code review.
In CodeAgent, security analysis is treated as one of the code review subtasks.
The framework requires the LLM agent to evaluate the security of a commit based on a predefined set of 25 vulnerability factors proposed by the authors.
These factors capture common vulnerability patterns and risky coding practices, guiding the agent to assess whether a commit may introduce security risks.

\subsection{Evaluation Metrics}

Following prior work, we evaluate commit-level classification performance using Precision, Recall, and F1-score.
\section{Experimental Results}

\subsection{Overall Performance}

\begin{table}[!tbp]
    \centering
    \caption{Performance comparison.}
    \label{tab:rq1_results}
    \small
    \begin{tabular}{l l r r r}
        \toprule
        \textbf{LLM} & \textbf{Method} & \textbf{Precision} & \textbf{Recall} & \textbf{F1-score} \\
        \midrule
        \multirow{4}{*}{DeepSeek-V3.2}
         & Direct      & 67\% & 9\%  & 16\% \\
         & CoT         & \textbf{86\%} & 6\%  & 11\% \\
         & CodeAgent   & 64\% & 22\% & 33\% \\
         & \toolname   & 70\% & \textbf{48\%} & \textbf{57\%} \\
        \midrule
        \multirow{4}{*}{GPT-4o-mini}
         & Direct      & 18\% & 8\%  & 11\% \\
         & CoT         & \textbf{57\%} & 15\% & 24\% \\
         & CodeAgent   & 41\% & 60\% & 49\% \\
         & \toolname   & 50\% & \textbf{75\%} & \textbf{60\%} \\
        \midrule
        \multirow{4}{*}{Qwen-Plus}
         & Direct      & 75\% & 28\% & 41\% \\
         & CoT         & \textbf{77\%} & 19\% & 31\% \\
         & CodeAgent   & 52\% & \textbf{58\%} & 55\% \\
         & \toolname   & 76\% & \textbf{58\%} & \textbf{66\%} \\
        \bottomrule
    \end{tabular}
\end{table}

Table~\ref{tab:rq1_results} presents the performance comparison between \toolname and baseline methods across three LLMs.

\textbf{Substantial improvement in recall.}
Across all evaluated LLMs, \toolname consistently achieves significantly higher recall than Direct and CoT baselines.
This demonstrates that \toolname is substantially more effective in identifying VICs, which is critical in security scenarios where missed vulnerabilities can lead to severe consequences~\cite{10172583}.

\textbf{Limitations of Direct and CoT baselines.}
Direct and CoT generally achieve high precision but extremely low recall.
For instance, CoT with DeepSeek-V3.2 achieves a precision of 86\% but only 6\% recall indicating that these methods are overly conservative.

\textbf{Comparison with CodeAgent.}
Although CodeAgent improves recall compared to Direct and CoT, it remains inferior to \toolname across all LLMs.
For example, under DeepSeek-V3.2, CodeAgent achieves an F1-score of 33\%, whereas \toolname reaches 57\%.
Similar trends are observed for GPT-4o-mini and QWen-Plus.
These results suggest that simply introducing agent-based reasoning is insufficient; instead, the structured multi-stage design of \toolname plays a crucial role in improving effectiveness.



\subsection{Ablation Study}
\label{sec:ablation}

We conduct an ablation study using DeepSeek-V3.2 to evaluate the contribution of each component in \toolname.
Specifically, we evaluate the following variants:

    
    
    

\textbf{w/o Code Analyst (w/o CA):} The CA is removed, and the VI directly analyzes the commit without structural information extracted from code changes.

\textbf{w/o Target Analyst (w/o TA):} The TA is removed, and the VI performs vulnerability analysis without explicit guidance on commit intent and goal-file alignment.

\textbf{w/o Stage 2:} The reanalysis stage is removed, and the VI makes final decisions solely based on the preliminary inspection without refinement.

\textbf{w/o Stage 3:} The final decision stage is removed.
The refined results from Stage 2 are directly used as the final prediction, without the additional fragment-level verification and commit-level aggregation performed in Stage 3.

\begin{table}[!tbp]
    \centering
    \caption{Ablation results of \toolname.}
    \label{tab:ablation}
    \small
    \begin{tabular}{lccc}
    \toprule
        \textbf{Method} & \textbf{Precision} & \textbf{Recall} & \textbf{F1-score} \\
    \midrule
        w/o CA & 65\% & 35\% & 46\% \\
        w/o TA & 64\% & 45\% & 53\% \\
        w/o Stage 2 & \textbf{72\%} & 43\% & 54\% \\
        w/o Stage 3 & 67\% & \textbf{57\%} & \textbf{62\%} \\
        \toolname & 70\% & 48\% & 57\% \\
    \bottomrule
    \end{tabular}
\end{table}

\begin{table}[!tbp]
    \centering
    \caption{Confusion matrix comparison of ablation variants.}
    \label{tab:confusion_final}
    \small
    \begin{tabular}{lcccc}
        \toprule
        \textbf{Method} & \textbf{TP} & \textbf{FP} & \textbf{T}N & \textbf{FN} \\
        \midrule
        w/o CA & 37 & 20 & 115 & 69 \\
        w/o TA & 48 & 24 & 111 & 58 \\
        w/o Stage 2 & 46 & 18 & 117 & 60 \\
        w/o Stage 3 & 60 & 30 & 105 & 46 \\
        \toolname & 51 & 22 & 113 & 55 \\
        \bottomrule
    \end{tabular}
\end{table}

These variants allow us to evaluate the impact of structural context (CA), intent understanding (TA), Stage 2 reanalysis, and Stage 3 final decision on the overall performance of \toolname.
Table~\ref{tab:ablation} and Table~\ref{tab:confusion_final} present the results of the ablation study.

\textbf{Impact of Stage 2 reanalysis.}
Removing Stage 2 increase precision but reduces recall and F1-score.
The confusion matrix shows a decrease in TP (51→46) and an increase in FN (55→60), indicating that more vulnerabilities are missed.
This suggests that Stage 2 plays an important role in improving coverage.
Given candidate fragments and their predicted vulnerability categories, the model can perform more targeted and type-guided reasoning under a narrowed search space.
This refinement step primarily serves to re-examine and disambiguate the suspicious fragments identified in Stage 1, allowing the model to make more informed decisions on borderline or uncertain cases.
As a result, more potential vulnerabilities are retained as positive predictions, which improves recall but may also introduce additional false positives.
Therefore, Stage 2 is a key component in the coarse-to-fine reasoning process, improving coverage while introducing a moderate trade-off between recall and precision, and contributing to better overall detection effectiveness.

\textbf{Impact of Stage 3 final decision.}
Removing Stage 3 increase recall and F1-score.
However, the confusion matrix shows that this gain is accompanied by lower prediction reliability.
Specifically, when Stage 3 is removed, FP increases from 22 to 30, while TN decreases from 113 to 105.
This provides direct evidence that Stage 3 filters out a subset of false positives that remain after Stage 2.
At the same time, TP decreases from 60 to 51 and FN increases from 46 to 55 after Stage 3 is enabled, showing that the additional verification step also filters out some true positives.
Therefore, Stage 3 introduces a clear trade-off: it sacrifices some recall in exchange for stricter confirmation and improved robustness.
Overall, these results show that Stage 3 acts as a conservative filtering mechanism over the refined candidates produced by Stage 2.
Its primary role is not to maximize F1-score, but to improve the trustworthiness of final predictions by reducing unreliable positive decisions.
In practical vulnerability detection scenarios, this behavior can be desirable because excessive false positives may lead to unnecessary security alerts and increased manual inspection effort.



\subsection{Efficiency and Cost Analysis}

\begin{table}[!tbp]
    \centering
    \caption{Average cost per commit evaluated with DeepSeek-V3.2.}
    \label{tab:cost}
    \small
    \begin{tabular}{l r r}
        \toprule
        \textbf{Method} & \textbf{Query Time (s)} & \textbf{Cost in USD} \\
        \midrule
        Direct & 5 & 0.0079 \\
        Cot & 17 & 0.0083 \\
        CodeAgent & 80 & 0.0217 \\
        \toolname & 103 & 0.0487 \\
        \bottomrule
    \end{tabular}
\end{table}

We evaluate the computational efficiency of \toolname by comparing its query latency and monetary cost with baseline methods.
Table~\ref{tab:cost} reports the average cost per commit using DeepSeek-V3.2.
The reported cost is calculated based on the cache miss pricing, which provides a conservative upper-bound estimate of the actual cost.

\textbf{Higher cost due to multi-stage reasoning.}
Compared to Direct and CoT, which require only a single LLM call, \toolname incurs a higher cost due to multiple agent interactions across different reasoning stages.
This increase is expected, as \toolname performs structured multi-stage reasoning instead of single-pass inference.

\textbf{Cost-effectiveness compared to CodeAgent.}
Compared with CodeAgent, \toolname exhibits a moderate increase in cost, while achieving substantially higher recall and F1-score.
This suggests that the additional cost is justified by improved detection effectiveness.
Although \toolname introduces additional LLM calls across multiple stages, the design is suitable for offline or CI-integrated security auditing scenarios, where accuracy is prioritized over latency.
The modular structure also allows early-stage filtering to reduce unnecessary downstream analysis.


\subsection{Generalization on Recent Vulnerabilities.}
\label{RQ4}
To evaluate the generalization ability of \toolname under realistic conditions, we conduct experiments on a set of commits associated with recently disclosed vulnerabilities.
The evaluation dataset consist of 20 commits collected from 11 CVEs, including 11 VICs and 9 VFCs.
Importantly, the selected CVEs are published between March 2025 and March 2026, with the majority (10 CVEs) released after November 2025.
These vulnerabilities are likely to be beyond the training cutoff of DeepSeek-V3.2, making it unlikely that the model has prior exposure to these specific cases.
Table~\ref{tab:no_leakage} presents the performance of different methods on commits associated with recently disclosed vulnerabilities using DeepSeek-V3.2.

\begin{table}[!tbp]
    \centering
    \caption{Performance on recent CVEs (DeepSeek-V3.2).}
    \label{tab:no_leakage}
    \small
    \begin{tabular}{lccc}
        \toprule
        \textbf{Method} & \textbf{Precision} & \textbf{Recall} & \textbf{F1-score} \\
        \midrule
        Direct    & 80\% & 36\% & 50\% \\
        CoT       & \textbf{100\%} & 18\% & 31\% \\
        CodeAgent & 63\% & 45\% & 53\% \\
        \toolname & 67\% & \textbf{55\%} & \textbf{60\%} \\
        \bottomrule
    \end{tabular}
\end{table}

\textbf{Strong generalization to unseen vulnerabilities.}
\toolname achieves the best overall performance, with the highest recall, outperforming all baselines.
This indicates that \toolname is more effective in identifying VICs even when vulnerabilities are newly disclosed.


\textbf{Effectiveness of multi-stage reasoning.}
Compared to CodeAgent, \toolname improves recall and F1-score, demonstrating the benefit of the proposed multi-stage reasoning process.
This suggests that structured candidate generation and refinement are essential for detecting vulnerabilities in previously unseen scenarios.
\section{Discussion}

\textbf{Effect of Multi-Stage Reasoning.}
The multi-stage reasoning process follows a coarse-to-fine paradigm. The preliminary inspection stage prioritizes recall by generating candidate vulnerability hypotheses, while the reanalysis stage improves reliability through type-guided verification. Together, these stages balance exploration and verification, contributing to improved detection performance.



\textbf{Limitations.}
Despite its effectiveness, \toolname has several limitations.
First, it relies on LLM reasoning, which may still produce inconsistent results in complex scenarios.
Second, the effectiveness of the reanalysis stage depends on the availability and relevance of retrieved cases.
Third, the final decision stage introduces a conservative bias that may reduce recall for subtle vulnerabilities.
Finally, the framework focuses on commit-level detection and does not address fine-grained tasks such as vulnerability localization or patch generation.
\section{Threats to Validity}

We discuss internal and external validity threats for \toolname. Internally, although the multi-stage design improves reliability, LLMs may produce inconsistent or incorrect judgments. Prompt design and temperature settings can also affect outcomes. Iterative refinement mitigates but does not fully eliminate errors. Externally, our evaluation on the V-SZZ dataset may not generalize to all projects, programming languages, or real-world environments, and performance could vary across different codebases or vulnerability patterns.




\section{Related Work}
\label{sec:related work}


\textbf{Just-In-Time Vulnerability Detection.}
Early JIT vulnerability detection approaches rely on handcrafted features and traditional machine learning models~\cite{pornprasit2021jitline,yang2017vuldigger,ni2022best,riom2021revisiting}.
More recent studies has explored deep learning techniques to learn representations from source code and commit diffs, including DeepJIT~\cite{hoang2019deepjit}, CodeJIT~\cite{nguyen2024code}, and HgtJIT~\cite{sun2025hgtjit}.
Although these approaches improve semantic modeling, they still struggle to capture developer intent and broader contextual information associated with commits.

\textbf{Code Vulnerability Detection.}
Code vulnerability detection has evolved from role-based and static-analysis approaches to learning-based methods.
Deep learning models such as VulDeePecker~\cite{li2018vuldeepecker} and VulDeeLocator~\cite{li2021vuldeelocator} learn vulnerability patterns from code representations, while graph-based approaches (e.g., Devign~\cite{zhou2019devign}, BGNN4VD~\cite{cao2021bgnn4vd}, COCA~\cite{cao2024coca}) model structural dependencies in programs.
More recently, transformer-based models such as LineVul~\cite{fu2022linevul} leverage pretrained language models for vulnerability prediction.
However, these methods primarily focus on static code snippets and are not designed for commit-level vulnerability reasoning.

\textbf{LLM-based Vulnerability Analysis.}
Recent advances in LLMs have demonstrated strong capabilities in code understanding and vulnerability analysis~\cite{wang2025contemporary,jiang2026survey,li2025prompting,lu2024grace,zhou2025large,lin2025large}.
Existing approaches include prompt-based methods, retrieval-augmented methods~\cite{du2024vul}, and agent-based framework such as CodeAgent~\cite{tang2024codeagent}.

ReAct-style agents~\cite{yildiz2025benchmarking} further introduce iterative reasoning through thought--action--observation loops.
However, existing LLM-based approaches primarily focus on source-code vulnerability analysis or general code review, and often rely on single-agent or single-pass reasoning.
As summarized by recent surveys~\cite{zhou2025large}, current approaches remain sensitive to prompt design and may produce unstable reasoning results. These limitations motivate our work, which focuses on commit-level vulnerability detection and employs structured role decomposition together with multi-stage verification.
\section{Conclusion}
\label{sec:conclusion}

In this paper, we propose \toolname, an LLM-based multi-agent framework for detecting vulnerability-inducing commits.
The framework formulates vulnerability detection as a structured multi-stage reasoning process that integrates code diffs, commit messages, and contextual information.
Experimental results show that \toolname consistently outperforms baseline methods across multiple LLMs, achieving up to 2$\times$ higher recall and 1.2--1.7$\times$ higher F1-scores compared to the strongest baseline.
These results demonstrate the effectiveness of coarse-to-fine reasoning for commit-level vulnerability detection, enabling improved coverage while maintaining reliable decision making.
In future work, we plan to explore more adaptive decision strategies and extend the framework to support fine-grained vulnerability classification.

\begin{credits}
\subsubsection{\ackname} This work was supported by the National Key R\&D Program of China No 2024YFB4506200.
\end{credits}

\bibliographystyle{splncs04}
\bibliography{reference}

\end{document}